\documentclass[useAMS,usenatbib]{mn2e}
\usepackage{graphicx}
\usepackage{txfonts}
\usepackage{fixltx2e}
\usepackage{multirow, array} 
\usepackage{float} 
\usepackage{booktabs}
 \def\apj {ApJ}
 \def\apjl {ApJL}

 \def\aap {A\&A}
 \def\mnras {MNRAS}

\title[Sensitivity of extrasolar mass--loss rate ranges]{On the sensitivity of extrasolar mass--loss rate ranges:\\ 
HD 209458b a case study}
\author[Villarreal D'Angelo et al.]
{C. Villarreal D'Angelo$^{1}$,  M. Schneiter$^{1,2}$, A. Costa$^{1,2}$,
P. Vel\'azquez$^{3}$, A.  Raga$^{3}$, A. Esquivel$^{3}$ \\
$^{1}$Instituto de Astronom\'\i a Te\'orica y Experimental, IATE (CONICET and Universidad Nacional de C\'ordoba),   C\'ordoba, Argentina\\ 
$^{2}$Faculty of Ciencias Exactas, F\'\i sicas y Naturales Universidad Nacional de C\'ordoba, Argentina\\
$^{3}$Instituto de Ciencias Nucleares, Universidad Nacional Aut\'onoma de M\'exico, DF, Mexico}

\begin{document}

\date{Accepted 2013 November 28. Received 2013 November 12; in original form 2013 April 2013 }


\maketitle

\label{firstpage} 
 
\begin{abstract}

We present a 3D hydrodynamic study of the effects that different
stellar wind conditions and planetary wind structures have on the
calculated Ly$\alpha$ absorptions produced by a cometary tail during
transit. We concentrate, as a case study, on the known HD 209458b
case.

Initially, we assume a broad range of possible planetary mass-loss rate values: $\dot M_p =[1-7] \times 10^{10}$g
s$^{-1}$. Then, by comparing  the observational Ly$\alpha$ absorption with  the numerically derived  ones, 
we could constrain the $\dot M_p $ values within the given range.
We find that the planetary mass-loss rate does not
change dramatically for large changes in stellar wind speeds
$\sim[250-800]$ km s$^{-1}$ and temperature $\sim [3-7] \times
10^{6}$ K while keeping fixed the stellar mass-loss rate 
( $\dot{M}_{\star}=9.0 \times 10^{-14}M_{\odot}$ yr$^{-1}$).
The $\dot M_p$ range found is $\sim[3-5] \times 10^{10}$g
s$^{-1}$, depending upon the efficiency of the stellar wind to
transport heat (polytropic index $\Gamma\sim[1.01-1.13]$), leading to
different stellar wind speeds. Several models with anisotropic
evaporation profiles for the planetary escaping atmosphere were
carried out, showing that both, the escape through polar regions,
resembling the emission associated with reconnection processes, and
through the night side, produced by a strong stellar wind that
compresses the planetary atmosphere and inhibits its escape from the
day hemisphere yields larger absorptions than an
isotropic planetary wind.

\end{abstract}


\section{Introduction}
The giant hot Jupiter HD 209458b was the first extrasolar planet
observed in transit \citep{char00,henry00} and the first for which
far-ultraviolet space observations were obtained with the \textit{Hubble
Space Telescope/}Space Telescope Imaging Spectrograph (STIS) instrument.\\ 
An initial analysis of this data set was performed by
\citeauthor{vidal03} (2003, hereafter VM03).  These authors carried out
observations in the Ly$\alpha$ line (at 1215.67 \AA), measuring an
extra mid-transit absorption (relative to the level of the continuum)
of $15 \pm 4$ per cent over a range of [-130,200] km s$^{-1}$.  Also, the
spectral absorption during transit was deeper on the blue side of the
stellar line.  These results lead the authors to claim the existence
of an extended exosphere (as predicted in \citealp{sch98}). Later, a
new analysis of the same data set was performed by \citeauthor{bj2007}
(2007, 2008, hereafter BJ07 and BJ08, respectively) to account for the
effect of stellar variability on the transit depths. In these works,
the authors measured an absorption of $8.9 \pm 2.1$ per cent in the $\pm
200$ km s$^{-1}$ velocity range, which they explained as the
effective absorption made by a cloud of $\sim 2.7 R_p$ (which is
smaller than the planetary Roche lobe), implying no direct evidence
for a cometary tail in the data.\\
These differences in the
measurement of the Ly$\alpha$ absorption led to a discussion in the
literature (see BJ07; BJ08; \citealt{vm2008}). BJ07
developed a new correction for the geocoronal contamination that led
him to redefine the wavelength limits used for calculating the transit
depth.  Unlike BJ07, in \citeauthor{vm2008} (2008, hereafter VM08) 
the authors argued that
the discrepancies were mainly due to the use of different wavelength
(or velocity) ranges to convert the spectra, as a function of time,
into a single absorption depth, and to the reference flux used to
correct the intrinsic stellar flux variation.  Hence, they claimed
that larger wavelength domains imply a diluted absorption signal with
a consequently lower absorption depth. \citet{kosk} argued that the
apparent disagreement between BJ07 and VM03 arises not only from
different definitions of the transit depth but also from differences
in the treatment of the data.  However, they agreed with VM08 that, as
the core of the absorption line is optically thick up to the Roche
lobe, the atmosphere must be evaporating.\\ Another issue related to
the transit absorption of Ly$\alpha$ photons in the velocity range of
[-130,100] km s$^{-1}$, is the origin of the high-velocity neutral
gas.  To answer this question, and arguing that the stellar radiation
pressure is not enough to explain the observations, \citet{holm}
proposed that the high-velocity neutral hydrogen can be produced by
charge exchange between the neutral atmosphere and the stellar wind
ions. The stellar wind velocity they used was somehow low ($50$ km
s$^{-1}$) in comparison to the typical solar wind
conditions. Following this idea, and employing a more typical stellar
wind velocity ($450$ km s$^{-1}$), \citet{eken} modelled the production
of neutral hydrogen and matched the signature of the Ly$\alpha$
absorption profile. \\
Also, other atmosphere mass-loss models due to
EUV input from  host stars can also account for the Ly$\alpha$
absorption and the high-velocity neutral gas (see VM03;
\citealp{lecavelier2004}; \citealp{yelle2004}; \citealp{tian};
\citealp{gar07}; \citealp{sch07}; \citealp{murr}; \citealp{kosk};
 \citealp{guo}). \\ 
The proximity of the planet to its parent star
($\simeq 0.045$ au) and the consequent non-thermal heating of the
upper planetary atmosphere (up to $10^{4}$K) by stellar X-ray and EUV
photons, would produce a hydrodynamic blow-off of the upper
atmospheric layers.  \citet{lecavelier2004} showed that this mechanism
may be enhanced by tidal forces that could pull the exobase level up
to the Roche limit, and predicted an asymmetric shape for the transit
curve due to the comet-like tail produced by the evaporating
atmosphere of the planet. This scenario was reinforced by the
detection of hydrodynamic escape signals in C II and O I lines (VM08),
Na I D lines \citep{vm2010} and Si III line \citep{linsky2010}.
Though, according to \citet{ehr08} the observations are not yet
sufficient to establish the presence of a tail.  \\The question
whether the STIS observations are correctly explained as a planetary
wind driven by photoionization or by charge exchange (or both
processes) remains open. The works of \citet{bj-h} and \citet{guo}
showed that either energetic neutral HI of stellar origin or thermal
HI populations in the planetary atmosphere could fit the Ly$\alpha$
observations. \citet{kosk} introduced a generic method to analyse and
interpret ultraviolet transit light curves. They applied it to the measurements
of HD 209458b, and showed that the observations can be explained by
absorption in the upper atmosphere without the need of ENAs.

Almost all models that try to explain the observed Ly$\alpha$
absorption of HD209458b produce mass-loss rates that are several times
$10^{10}$ g s$^{-1}$. For example, \citeauthor{yelle2004} (2004, see correction
in \citealt{yelle2006}), find ${\dot M}_p = 4.7 \times 10^{10}$ g
s$^{-1}$, \citet{tian} find ${\dot M}_p < 6 \times 10^{10}$ g s$^{-1}$,
\citet{gar07} finds ${\dot M}_p = 6.1 \times 10^{10}$ g s$^{-1}$,
\citet{guo} finds ${\dot M}_p = 3.4 \times 10^{10}$ g s$^{-1}$, and
\citet{murr} find ${\dot M}_p \sim 2 \times 10^{10}$ g s$^{-1}$.\\ 
We need to keep in mind that the presence of a planetary magnetic field can
lead to significant changes in the upper atmosphere of planets and
consequently can lower the planetary mass-loss rate.  \citet{adams},
considering an outflow modulated by a magnetic field, estimated a
planetary outflow of the order of $\sim 10^9$ g s$^{-1}$ produced
mainly in the polar regions.\\ 
In \citealt{sch07} (2007, hereafter SCH07) a model for
the cometary exosphere around HD 209458b was presented via the
assumption that it is dynamically similar to the ion tail of a comet
\citep[see e.g.][]{rau97}. 3D hydrodynamic simulations of the
interaction between the material, which is being photoevaporated from
the planetary atmosphere, and the impinging stellar wind lead to the
conclusion that a cometary structure, obtained as a consequence of the
simulated wind interaction, must have a size that grows for increasing
values of the planetary mass-loss rate ${\dot M}_p$, and therefore
produces a Ly$\alpha$ absorption of the stellar emission which is a
monotonically increasing function of ${\dot M}_p$. In this model, a
fixed solar-like value for the stellar wind speed and an isotropic
emission were assumed. From the comparison with the observations of
VM03 for the Ly$\alpha$ absorption, the authors estimated the 
mass-loss rate as ${\dot M}_p\sim 10^{10}$g s$^{-1}$.\\ 
In this work, despite the mentioned discussions in literature (not yet
settled), and assuming the existence of an evaporating atmosphere, we
study the influence of different stellar wind temperatures and
velocities, on the planetary mass-loss rate. In spite of the
oversimplifications of our model a variation of the stellar wind speed
and temperature, in a reasonable range, gives different values of
$\dot M_p$. Thus, different combinations of $(v_0, \dot M_p)$ 
($v_0$ is the stellar wind speed at the planet position) could
give account of the absorption derived from the observations. We
explore the allowed range of this combination of parameters and
discuss its implication on the determined stellar-planetary
parameters.
The procedure consists of assuming values of $\dot M_p$ within a range of possible ones.
Then, to constrain the planetary mass-loss rate the numerically derived  Ly$\alpha$ absorption is compared with the observations.\\
The first part of the study is performed assuming an
isotropic planetary wind for a thermalized atmosphere. In the second
part, we explore the effect of different structures of planetary wind
emissions resembling different cases such as the day/night asymmetry
due to a tidally locked planetary motion, and polar winds due to
emission associated with reconnection processes (see Fig.
\ref{fig1}). We emphasize that this final part only presents toy
atmospheric models which illustrate how more detailed escaping
atmosphere models could modify the predicted Ly$\alpha$
absorption.\\ We describe our numerical approach in \S \ref{method},
the anisotropic planetary ejection wind models in \S \ref{toy}, and
discuss the consequences of the results in \S
\ref{discussion}. Finally, we summarize the results and conclusions in
\S \ref{conclusion}.\\
 
\section{The method}
\label{method}

\textit{The simulation:} The 3D hydrodynamic simulations were carried
out with the Yguaz\'u code \citep{raga00} as described in
SCH07\nocite{sch07}   that evolves the ideal HD equations given, in conservative form, by
\begin{equation}
\frac{\partial \rho}{\partial t}+\mathbf{\nabla}\cdot(\rho\mathbf{u})=0\label{1}
\end{equation}
\begin{equation}
\frac{\partial \rho \mathbf{u}}{\partial t}+\mathbf{\nabla}\cdot(\rho\mathbf{u}\mathbf{u}+p \mathbf{I})=\rho\mathbf{g}\label{2}
\end{equation}
and
\begin{equation}
\frac{\partial \varepsilon}{\partial t}+\mathbf{\nabla}\cdot(\mathbf{u}(\varepsilon+p)=\rho\mathbf{g}\cdot\mathbf{u},\label{3}
\end{equation}
where $\rho$ indicates the density; {\bf \textit{u}} the velocity; {\bf \textit{g}} is the gravity,  \textit{p}
is the pressure and $\varepsilon$ is the total  energy density given by  
\begin{equation}
\varepsilon= \frac{p}{\gamma -1}+\frac{1}{2}\rho u^{2},
\label{4}
\end{equation}
where $\gamma$ is the ratio of specific heats. For the simulations we adopt $\gamma=\Gamma$ with $\Gamma$ being the polytropic index.

A six level, binary adaptive grid with a maximum resolution of $1.8
\times 10^4$~km was used.  A computational domain of $7.4\times
10^{7}$~km (or 0.49 au.), $1.85\times 10^{7}$ and $7.4\times
10^{7}$~km (in the $x$-, $y$- and $z$-directions, respectively) was
employed.  The coordinate system is such that the star is placed at
the centre of the domain and the planet is orbiting in the
$xz$-plane. We define two inner boundaries, one that belongs to the
planet, whose location is updated at each time step according to the
orbital motion, forced to have the maximum resolution, and the other
is located at a fixed radius from the center of the computational
domain, where the stellar wind is imposed continuously.

The gravity of both, planet and star were included in the
simulations. The radiation pressure was considered by reducing the
effective gravity as in VM03, i.e. the Ly${\alpha}$ radiation
pressure is assumed to be $0.7$ times the stellar gravity.

\noindent
\textit{The stellar wind:} The G0 V star (HD 209458) of $1.148 M_\odot$  was simulated to have a
 non magnetized isotropic wind, with mass-loss rate of
  $\dot{M}_{\star}=9.0 \times 10^{-14}M_{\odot}$~yr$^{-1}$
(corresponding to an ion flux of $5.2\times 10^{36}$ s$^{-1}$) fixed for all the models.  The
stellar inner boundary, where the wind is imposed continuously, is
launched at $R=6.9{R}_{\star}$.  For all the models used, this position
is located beyond the critical point where the wind velocity is still
increasing, before the terminal speed is reached. We used several
polytropic models to calculate reasonable initial conditions of the
almost thermally driven wind (with values of the polytropic index close to
$1$).  In this way, we could vary the velocity in the range
$\sim$[$200-800$] km s$^{-1}$, at the radius where the stellar wind is
imposed. These results are in agreement with the wind velocity
profiles in \citet{vido12} and shown in Table \ref{tab1}. In the
table, $T_0$, is the coronal temperature, $T_w$ and $v_w$ are,
respectively, the temperature and velocity at the position where the
wind is launched, calculated using two polytropic wind models (with
$\Gamma$ being the polytropic index $\Gamma=[0.01-1.13]$).  The density
varies consistently with the fixed value of $\dot{M}_{\star}=4\pi \rho
v R_{\star}^{2}$  ($R_{\star}=7.97 \times 10^5$ km being the stellar radius). At the
position of the planet the density values ($[1.2 - 4.1]\times  10^{-20}$gr cm$^{-3}$) are consistent with the
values found by other authors  (e.g. \citealp{vido10},
\citealp{llam11}).
 \begin{table}
 \centering
\begin{tabular}{c c c c c c| c c c c }

 \toprule

\multirow{1}{5mm}{$\Gamma$ } & {$T_0 $  MK}& 
                                      \multicolumn{4}{c}{$T_w $ MK} & \multicolumn{4}{c}{$v_w$ \ (km s$^{-1}$)}\\
\midrule
\multirow{5}{5mm}{1.01}     & \ \ \    3  \ \ \    & 2.8  & \ \ \   \ \ \ \ \  \ \ \ \ \ 372\\
			     & \ \ \    4  \ \ \   & 3.8   & \ \ \    \ \ \ \ \ \ \ \ \ \ 488\\
			     &  \ \ \   5 \ \ \    & 4.7  & \ \ \   \ \ \ \ \  \ \ \ \ \ 594 \\
			     & \ \ \    6  \ \ \   & 5.7  & \ \ \   \ \ \ \ \  \ \ \ \ \ 692\\
			     &  \ \ \   7  \ \ \   & 6.7  & \ \ \   \ \ \ \ \ \ \ \ \ \ 784\\
\midrule		     
\multirow{4}{5mm}{1.13}       & \ \ \   3  \ \ \      & 1.3 & \ \ \  \ \ \ \ \ \ \ \ \ \ 205  \\
			     & \ \ \    4   \ \ \    & 2.0  & \ \ \   \ \ \ \ \ \ \ \ \ \  349 \\
			     & \ \ \    5   \ \ \   & 2.6 & \ \ \   \ \ \ \ \ \ \ \ \ \  475 \\
			     &  \ \ \   6   \ \ \   & 3.2 & \ \ \   \ \ \ \ \  \ \ \ \ \ 590 \\

\bottomrule
\end{tabular}
\caption{Determination of the stellar wind inner boundary parameters.}
\label{tab1}
\end{table}

\noindent
\textit{The planet:} It was initially  modelled to have an isotropic
evaporating wind, emitted from $3 R_p$ with a velocity $v_p=10$
km s$^{-1}$ (where $R_{p}= 1.4 R_{Jup}$ and $M_{p}=0.69 M_{Jup}$). The
boundary condition for the base plasma temperature was fixed at
T$_{0}=10^{4}$ K and we varied the mass-loss rate between [1-7]
$\times 10^{10}$g sec$^{-1}$ corresponding to number densities
between $[9.4 x 10^{6}-6.6x10^{7}]$ cm$^{-3}$ at the base of the inner
boundary.  It is important to emphasize that we do not solve the
problem of the production of a wind from the irradiated surface of the
planet (for works that handle the wind production see;
e.g. \citet{preu} and \citet{tian}). Instead, we assume that the
planet ejects an (initially) isotropic wind from its upper atmosphere
with parameters that are consistent with the works of \citet{tian},
\citet{kosk}, \citet{murr} and \citet{guo}. 
The density profile inside the inner boundary ($r<3R_p$) is slightly
different to the one proposed in \cite{tian}, since it has 
a slower increase towards the surface of the planet. 
This does not affect the calculated absorption, since the material
within the Roche lobe ($< \sim 3 R_p$) absorbs in the range cover 
by the $Ly_\alpha$ geocoronal contamination.
Assuming an escaping planetary wind is a way of eliminating from our
simulations the complex problem of the calculation of the
photoevaporation of the planetary atmosphere.

In addition four models with an anisotropic planetary wind were run,
where the emitting surface was changed, and the wind velocities were
modified such that the mass-loss rates were conserved. These models
are thoroughly explained in Section \S \ref{toy}.

\noindent
\textit{Computation of the Ly$\alpha$ absorption:} From the computed
density, velocity and temperature structures we calculated the
Ly$\alpha$ absorption associated with a planetary transit of the stellar
disc.  The computational grid is oriented at an angle
$i=86^{\circ}.67,$ i.e. the angle between the orbital axis and the
line of sight of HD 209458b. The optical depth is obtained as

\begin{equation}
\tau_{\nu}= \int n_{HI} \sigma_{0} \phi (\Delta \nu)ds \,,
\label{tau}
\end{equation}

\noindent for all lines of sight, where the integral goes from the
surface of the star  to the edge of the computational domain, $n_{HI}$ is the
neutral hydrogen number density, $\nu$ is the frequency and $\Delta
\nu=\nu-\nu_0$ is the offset from the line centre,
$\sigma_{0}=0.01105$~cm$^2$ \citep{oster} is the Ly$\alpha$ absorption
cross-section at the line centre, and $s$ is the length measured along
the line of sight. $\phi (\Delta \nu)$ is the Doppler line profile given by

\begin{equation}
  \phi (\Delta \nu)=\Big(\frac{m_H}{2 \pi k T} \Big)^{1/2}
e^{-m_H {\Delta u}^2 /2kT} \frac{c}{\nu_{0}}\,,
\label{phi}
\end{equation}

\noindent where $m_H$ is the proton mass, $c$ is the velocity of
light, $k$ is the Boltzmann constant, $T$ is the gas temperature,
$\Delta u=u_{r}-u_F$, $u_F$ is the flow velocity along the line of
sight, and $u_r=c\Delta \nu/\nu_0$ is the radial velocity associated with
a frequency shift $\Delta \nu$ from the line centre $\nu_0$.

The stellar wind is assumed to be a fully ionized H gas so it does not
contribute to the absorption (i. e., $n_{HI}=0$ for equation
\ref{tau}). Also, the planetary wind is supposed to be fully neutral
[so that $n_{HI}={\rho}/(1.3m_H)$, where $\rho$ is the gas
density]. With the interaction the planetary material is heated and
mixed with the ionized stellar wind. We consider that the planetary
material temperature above its ionization value ($T\sim 10^{5}$K) will
not contribute to the absorption. The material from the planetary wind
is identified with an advected passive scalar in the simulations.

We consider that at a given frequency the stellar disc emits a uniform
specific intensity $I_{\nu,*}$ (i.e., we neglect the centre-to-limb
variation). We then compute the intensity seen by the observer
($I_\nu= I_{\nu,*} e^{-\tau_{\nu}}$, with $\tau_\nu$ given by
equation \ref{tau}). Finally, we carry out an integration of the intensity
over the stellar disc (for which we assume an $R_*=1.12R_{\odot}$
photospheric radius) in order to simulate an observation in which the
stellar disc is unresolved.

The total intensity decrease due to the absorption resulting from the
presence of the planetary exosphere was computed as a frequency
average to obtain

\begin{equation}
I/I_*={1\over{\nu_2-\nu_1}} \int^{\nu_2}_{\nu_1}<e^{-\tau_{\nu}}> d\nu\,,
\label{istar} 
\end{equation}

\noindent where $<e^{-\tau_{\nu}}>$ is the absorption averaged over the
stellar disc, and $\nu_1$ and $\nu_2$ correspond to the limits of
the velocity range.


\noindent
\textit{Choice of the absorption range:} The averaging of the
absorption over frequency was done, taking the ratio between the
emission (as a function of frequency) observed during transit and the
`stellar emission' (i. e., the emission observed away from transit),
and then integrating within a frequency band which includes the
Ly$\alpha$ line (for a more complete description see
SCH07). Hence, the values of the absorptions depend strongly
on the integration range utilized, and since several authors have
found different absorption values, due to the choice in wavelength
ranges that were employed, it is important to take them into account
when analysing the results.

For instance, VM03 reported an $\sim 15$ per cent Ly$\alpha$ absorption
integrating in the range $\left[-130,100 \right]$km s$^{-1}$, BJ07
reported a $\sim 8.9$ per cent absorption in the range $\left[-200,200
  \right]$km s$^{-1}$ and \citeauthor{vm2004} (2004, hereafter VM04) an $\sim 5$ per cent
Ly$\alpha$ absorption in the range $\left[-320,200 \right]
\mathrm{km\ s^{-1}}$.

With the purpose of comparing with the observations (VM03; VM04;
BJ07), to perform the Ly$\alpha$ integration we swept the three
mentioned velocity ranges, as shown in columns 4--6 in Table
\ref{tab2}.  The $\left[-130,100 \right]$km s$^{-1}$ was simulated to have
$\left[-120,100 \right]$km s$^{-1}$ in all models (this is due to the
discrete jumps in velocity, every $20$km s$^{-1}$, when calculating
the absorption). For all cases we took into account the geocoronal
contamination, performing the correction suggested in VM03, and
excluding the central part of the line, from $1215.5$ to
$1215.8$ \AA (i.e., $\left[-40,40 \right]$km s$^{-1}$).
To identify the absorption ranges we adopt the  convention used in VM08: 
line core interval (C:  [19-11] per cent) measured by 
VM03, intermediate line range (I:   [11-6.8] per cent) measured 
by BJ07 and whole line range (W:  [7-3] per cent) measured by VM04.

\section{Anisotropic planetary wind models}
\label{toy}
The rotational periods of `hot Jupiters' are likely synchronized with
their orbital periods, hence only one side of the planet is
irradiated. For this reason, it is not a bad idea to relax the
assumption of isotropic, and try to see how different structures of
planetary winds affect the measured absorptions. Also, going a little
further in this line of thought, and based on the discussion in
\citet{adams} where the probable existent planetary magnetic field is
simulated, we propose a scenario where the escaping material occurs in
the polar regions.  With this in mind we present four oversimplified
alternative inner boundaries for the planet, that we named as follows

\begin{figure*}
\centering
\includegraphics[width=16cm]{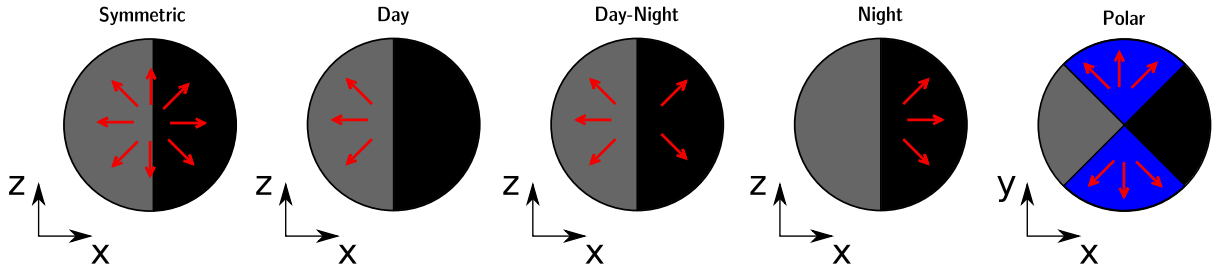}
  \caption{Schemes of inner boundary setups for the different
    planetary wind models corresponding to isotropic emission (first
    panel), only day emission (second panel), day/night asymmetry
    (third panel), only night emission (fourth panel) and polar
    emission (fifth panel).}
  \label{fig1}
\end{figure*}

\begin{itemize}

\item \textit{Day}: escape only through the illuminated hemisphere.  This
  is the case when the energy deposited in the day side cannot be
  redistributed to the night side, as would happen when the horizontal
  advection time is longer than the radial advection time (see
  discussion in \citealp{murr}).

\item  \textit{Day-Night}: 75 per cent of the mass escapes from the day side and
  25 \% from the night side. This happens because some of the energy
  deposited on the illuminated hemisphere is transported to the other
  hemisphere, but a variation remains \citep{yelle2004}.

\item \textit{Night}: escape only through the night side. A strong
  stellar wind may compress the atmosphere and inhibit its escape from
  the day hemisphere \citep{gar07}.

\item \textit{Polar}: escape through the poles resembling reconnection
  processes of a dipolar planetary magnetic field with the magnetic
  field carried out by the stellar wind \citep{adams}. In our model
  the wind is emitted from the polar surfaces with areas determined by
  the solid angle $2\pi(1-cos(\theta))$ sr, where $\theta$ is measured
  from the $z$ axis and has a value of $\pi/4$.

\end{itemize} 
  To simulate all the anisotropic models   we chose the stellar parameters  of
  one  isotropic model corresponding to 
    the coronal temperature
   $T_0=$ MK and polytropic index $\Gamma=1.13$. For the planet, the
   wind velocity is such that the mass-loss rate is the same for all
   models (${\dot M}_p= 3 \times 10^{10}$g s$^{-1}$),
   i.e. $V_p=2V_{p(isotropic)}$ for the Day and Night models, since
   the surface from which the wind is emitted is halved. For the
   Day-Night the velocity is distributed according to the mass-loss
   rate, i.e.  $V_p(Day)=1.5V_{p(isotropic)}$, and
   $V_p(Night)=0.5V_{p(isotropic)}$. Finally, the velocity
   distribution for the Polar model is
   $V_p(Polar)=3.4V_{p(isotropic)}$. The absorption was calculated in
   the same way as  the isotropic wind model.  Fig. \ref{fig1}
   displays a scheme of the `planetary wind' for each `anisotropic
   model'.

\section{Results \& Discussion}

\label{discussion}

\subsection{Isotropic planetary wind models}

In SCH07, typical solar wind conditions were employed to simulate the
interaction with the escaping planetary atmosphere, and to compute the
resulting absorption. The $\sim 500 \mathrm{km\ s^{-1}}$ stellar wind
velocity used in SCH07 led to an ${\dot M}_p\sim 1.1 \pm 0.2 \times
10^{10}\mathrm{g\ s^{-1}}$ estimate (obtained by adjusting the
computed absorption to the value of the absorption reported by VM04).

One of the aims of this work is to explore the effect of
several wind speeds $v_{o}$, at the orbital position and coronal
temperatures $T_{o}$ on the inferred ${\dot M}_p$.
 For each model, columns 1-3 of Table \ref{tab2} show the values of  $T_{o}$,  $v_{o}$ and ${\dot M}_p$, respectively.
The table, also displays 
the maximum transit absorption depths for
each model calculated for the spectral domains as defined by
VM04, BJ07 and VM03 (columns 4 to 6, C, I and W, respectively) and the transit
time ($\Delta t$) \footnote{We refer to the transit time as the
  interval while there is Ly$\alpha$ absorption}. 
  The results are
divided in two parts, the upper part, that corresponds to models with
the initial conditions obtained with $\Gamma=1.01$, and the lower one,
that corresponds to models with $\Gamma=1.13$.  Columns 4-6 show the
maximum absorption values when excluding the part of the spectrum
contaminated by the geocoronal Ly$\alpha$ emission. Note that the
calculated absorptions correspond only to the contribution of the
tail; the material close to the atmosphere has a radial speed of $\sim 10$ km s$^{-1}$
 which falls within the subtracted speed range.
 
When comparing among the models some general trends can be noted.
Larger absorption values are obtained when lowering the coronal temperature,
and/or the stellar wind velocity, or by increasing the planetary mass-loss
rate.  This last trend is easily explained since as more planetary
material is ejected a larger column density is obtained, and
consequently a larger absorption can be expected.  Higher coronal
temperatures produce an increase of the temperature gradient - between
the stellar wind and the neutral planetary atmosphere ($\sim
10^{4}$K) - causing a larger heat transfer which increases the
ionization rate, thus lowering the absorption.  Also, higher stellar wind
speeds (due to higher $T_0$ or lower $\Gamma$ values) result in a
consequent larger stellar ram pressures that hinders the expansion of the
planetary wind, reducing the absorption and transit time.
 
The dependence of the transit time on $T_0$ and $v_0$ obeys the same
mechanisms as the absorption. That is, as an increase in $T_{0}$ is
accompanied by an increase in both, the ionization rate and the stellar ram
pressure, it produces changes in the size and structure of the
escaping material. More specifically, a larger ram pressure produces a
more radially aligned wake, and a more effective ionization leads to a
smaller size of the absorbing tail, and hence to shorter transit
times.
 
 For $\Gamma=1.13$ and all $T_{0} $ the models with ${\dot M}_p=3
 \times 10^{10}$ g s$^{-1}$ (highlighted values in the lower panel of Table
 \ref{tab2}) are the ones that better adjust the observational values for the three 
 spectral domains, i.e., they fit the limits given by VM03  ([19-11] per cent), 
BJ07 ([11-6.8] per cent) and  VM04 ([7-3] per cent).
  Whereas, for the models with $\Gamma=1.01$ larger
 ${\dot M}_p$ are required (highlighted in the upper panel of Table
 \ref{tab2}).  $\Gamma=1.01$ implies a more efficient heat transfer,
 hence the resulting stellar wind velocity and temperature at the
 orbital position will be larger than for $\Gamma=1.13$ resulting in
 lower absorption values.
 
 \begin{figure*}
\centering
\includegraphics[width=19.cm]{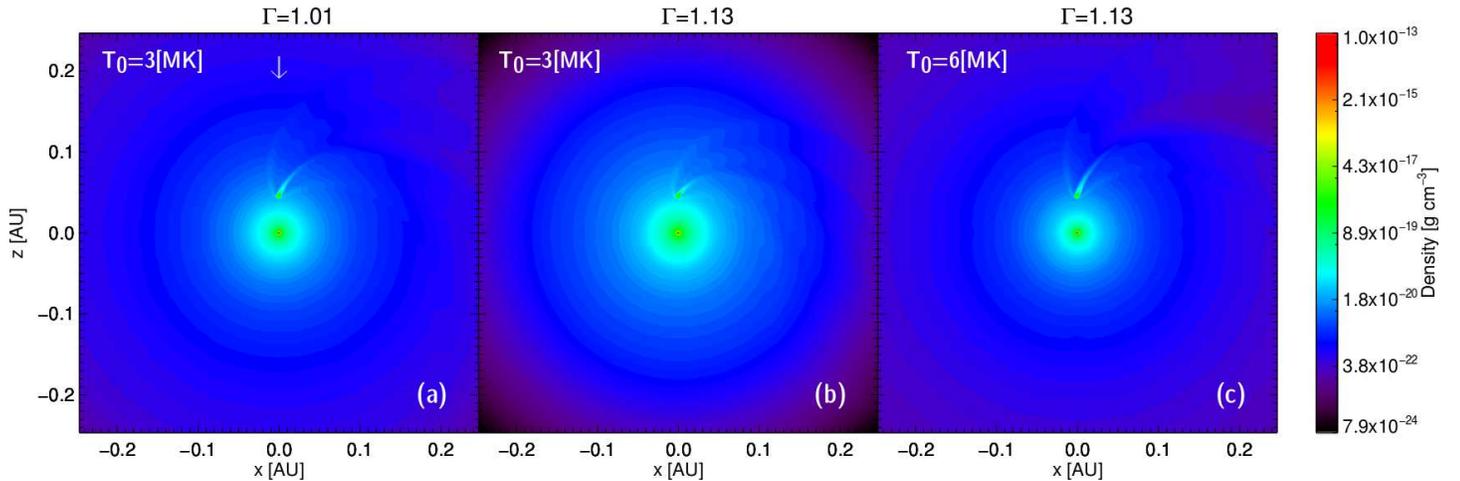}
   \caption{Orbital cuts of 3D HD simulation corresponding to the
     density stratification for a stellar wind temperature and
     polytropic index of (a) $T=3$ MK, $\Gamma=1.01$; (b)
     $T=3$ MK, $\Gamma=1.13$ and (c) $T=6$ MK,
     $\Gamma=1.13$. All figures are obtained when the planet is located at the same  orbital position.}
\label{fig2}
\end{figure*}

\begin{figure*}
\centering
\includegraphics[width=19.cm]{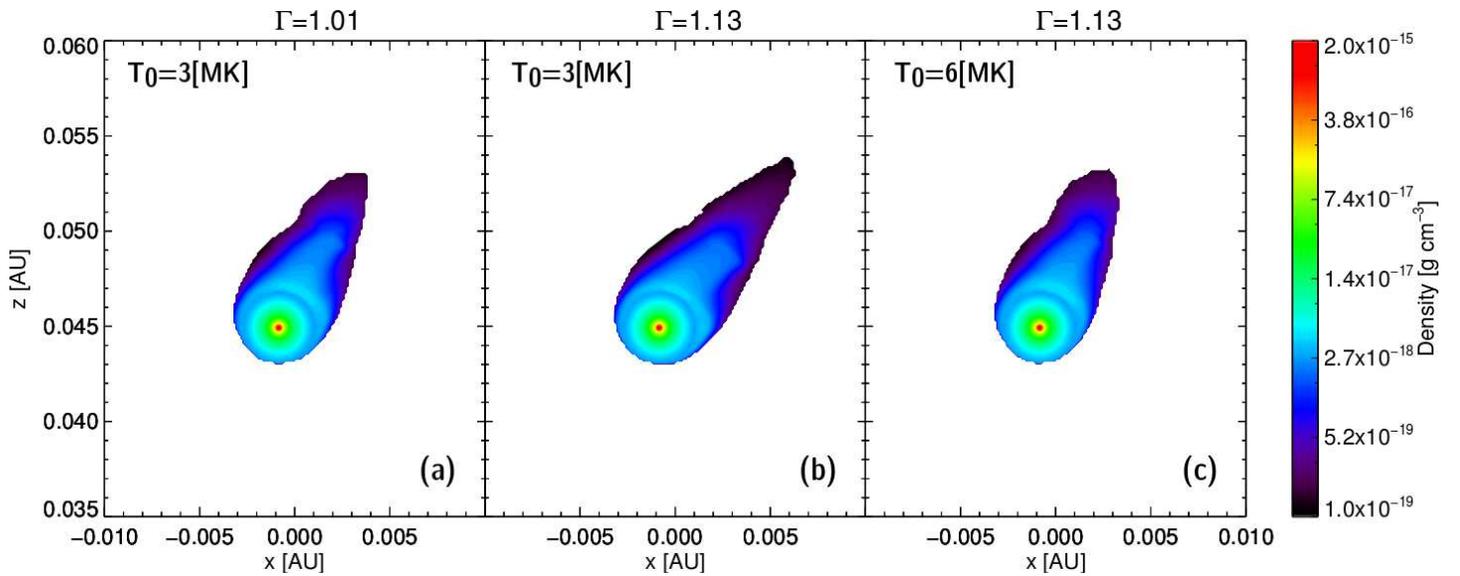}
   \caption{Enlarged orbital cuts of 3D HD simulation for ${\dot M}_p = 3 \times 10^{10}$ g s$^{-1}$
     corresponding to a density stratification of the absorbing
     planetary material for a stellar wind temperature of (a) $T=3$ MK, $\Gamma=1.01$; 
    (b) $T=3$ MK, $\Gamma=1.13$ and (c) $T=6$ MK, $\Gamma=1.13$. 
   All figures are obtained when the planet is located at the same  orbital position.  }
   \label{fig3}
\end{figure*}

Figs \ref{fig2}(a)-(c) display a comparison of the global density
distribution observed on the orbital plane for models with the same
${\dot M}_p=3\times 10^{10}$ g s$^{-1}$, noted in last column of Table \ref{tab2} as a, b and c,
respectively.  Figs \ref{fig2}(a) and (b) have the same
coronal temperature ($3$ MK) and different $v_o$ (or $\Gamma $)
($429~\mathrm{km\ s^{-1}}$ and $247~\mathrm{km\ s^{-1}}$,
respectively), while Figs \ref{fig2}(b) and (c) have the same
$\Gamma $ and different temperatures ($3$ and
$6$~MK, respectively).  The arrow in Fig. \ref{fig2}(a)
indicates the line-of-sight direction. Fig. \ref{fig2}(a) shows a
slightly more radially collimated tail than the ones in Fig.
\ref{fig2}(b) due to higher $v_o$.  When comparing Figs \ref{fig2}(b)
and (c) we see again that the comet-like tail for the larger
$T_{0}$ model becomes more oriented along the radial direction, as
noted earlier.  To resume, higher stellar wind speeds relative to the
orbital one ($v_{orb}\approx 146~\mathrm{km\ s^{-1}}$) imply
straighter tails.
  
 \begin{table*}
\centering
 \begin{tabular} {c c c c c c c}
 \toprule
$T_{0}$ (MK) &$v_o$ (km/s)&${\dot M}_p$ x$10^{10}$ (g/s) &C: [19-11] per cent  & I: [11-6.8] per cent &W:  [7-3] per cent& $\Delta t$ (h)\\
 & & & $\left[-120,100 \right]$km s$^{-1}$ &   $\left[-200,200 \right]$km s$^{-1}$&  $\left[-320,200 \right]$km s$^{-1}$&   \\
\midrule
\multirow{5}{5mm}{3}    &  \multirow{5}{5mm}{429}    &    1     & 6.30 & 2.76 & 2.01 & 4.4    \\
			                         &      &  \bf{  3   } & \bf{14.66} & \bf{6.43} &\bf{ 4.68 }& \bf{5.0} (a) \\
			                          &     &    5      & 21.51 & 9.72 & 7.05 & 5.5  \\
			                          &     &    7    & 26.09 & 12.61 & 9.17 & 6.1 \\
\midrule
\multirow{5}{5mm}{4}  &  \multirow{5}{5mm}{548} & 1        & 6.04 & 2.64 & 1.92 & 4.4  \\
		                              &    &3       &13.63  &5.97 & 4.34& 5.0 \\
			                      &    &\bf{5     }  & \bf{18.99 } & \bf{8.41} & \bf{6.12} & \bf{5.0}\\
		                              &    &    7    & 24.25 & 11.72 &8.53 & 5.5  \\
\midrule
\multirow{5}{5mm}{5}  &  \multirow{5}{5mm}{657}  &1     & 5.91 & 2.59 & 1.88 & 4.4 \\
			                      &     &3       & 12.78 & 5.60 & 4.07 & 5.0 \\
			                      &     &\bf{5      }& \bf{18.69 }& \bf{8.23 }&\bf{ 5.99 }&\bf{ 5.0} \\
			                       &    &    7   & 22.53 & 10.14 & 7.37 & 5.5  \\
\midrule
\multirow{5}{5mm}{6}   &  \multirow{5}{5mm}{757}  &1       & 5.67 & 2.48 & 1.80 & 4.4  \\
			                         &   &3      & 11.98 & 5.25  & 3.82 & 4.4 \\
			                         &   &\bf{5 }     &\bf{17.79}  &\bf{ 7.83 } & \bf{5.69} &\bf{ 5.0 }\\
			                          &   &    7    & 21.16 & 9.47 & 6.89 & 5.0  \\
\midrule
\multirow{5}{5mm}{7}   &  \multirow{5}{5mm}{852}  &1        & 5.49 & 2.40 & 1.75 & 4.4  \\
			                          &  &3     & 11.28 & 4.94  & 3.59 & 4.4 \\
			                          &  &\bf{5}       &\bf{16.92}  &\bf{ 7.42}  & \bf{5.39} &\bf{ 4.4} \\
			                         &   &    7    & 20.72 & 9.18 & 6.68 & 5.0  \\
\midrule
\midrule
\multirow{5}{5mm}{3}       &  \multirow{5}{5mm}{247} &  1     & 7.66 & 3.35 & 2.44 & 5.0    \\
						  &     &   \bf{3}     & \bf{17.44} & \bf{7.67} & \bf{5.58} & \bf{5.0} (b) \\
						  &     &    5 & 20.26 & 8.93 & 6.49 & 6.6  \\
						 &      &    7  & 22.82 & 10.10 & 7.35 & 7.2 \\
\midrule
\multirow{5}{5mm}{4}    &  \multirow{5}{5mm}{388} & 1       & 6.84 & 2.99 & 2.18 & 4.4  \\
		                                &    &\bf{3   }    & \bf{15.89 } & \bf{7.00} &\bf{ 5.09 }& \bf{5.5} \\
			                        &    &5       & 22.32  & 10.07 & 7.32 & 6.1\\
		                                  &  &    7   & 26.26 & 12.79 & 9.3 & 7.2  \\
\midrule
\multirow{5}{5mm}{5}  &  \multirow{5}{5mm}{513}  &1       & 6.35 & 2.78 & 2.02 & 4.4 \\
			 &  &\bf{3  } &  \bf{15.13} & \bf{6.64} & \bf{4.83} &\bf{ 5.0 }\\
			 &  &5   &   22.41 & 10.28 & 7.48 & 5.5 \\
			 & &7      & 27.10 & 13.99 & 10.17 & 6.7  \\
\midrule
\multirow{5}{5mm}{6}   &  \multirow{5}{5mm}{630}  &1      & 6.21 & 2.72 & 1.98 & 4.4  \\
			              &              &\bf{3 }  &  \bf{ 14.54} & \bf{6.38 } & \bf{4.64} &\bf{5.0} (c)\\
			                &            &5       &19.52  & 8.69  & 6.32 & 5.0 \\
			                 &            &    7    & 25.98 & 13.26 & 9.64 & 5.5  \\
\midrule
\bottomrule
\end{tabular}
\caption{
Isotropic planetary wind models. Parameters and results for  initial conditions with polytropic
  index upper: $\Gamma=1.01$ and bottom: $\Gamma=1.13$. Parameters: $T_{0}$ the
  coronal temperature, $v_{o}$ is the stellar wind speed at the planet
  position and ${\dot M}_p$ is the planetary mass-loss rate. Results: C 
  the line core interval comparable with  VM03 limits ([19-11] per cent); I 
  the intermediate line range comparable with  BJ07 limits ([11-6.8] per cent) and 
  W the whole line interval comparable with  VM04 limits ([7-3] per cent).
  $\Delta t$ is the transit time. Results in boldface  better adjust all the observational cases. 
  Models noted as a, b and c (last column) correspond to Figs \ref{fig2}(a)-(c), respectively.}
\label{tab2}
\end{table*}

\begin{table*}
\centering
 \begin{tabular} {c c c c c}
 \toprule
 Model & C: [19-11] per cent   & I: [11-6.8] per cent &W:  [7-3] per cent& $\Delta t$ [h] \\
\midrule
Polar      &   28.40 & 14.11 & 10.26 & 5.6 \\
Night      &   24.07 & 12.17& 8.85& 5.5 \\
Isotropic  &   15.89 & 7.00& 5.09& 5.5  \\
Day       &   15.53 & 6.85& 4.98 & 5.1 \\ 
Day-Night  &   12.76  & 5.59 & 4.07 & 5.0  \\
\bottomrule
\end{tabular}
\caption{Absorptions and transit times for the anisotropic planetary
  atmospheric models with  $\Gamma=1.13$, $T_0=4$ MK, and ${\dot M}_p=3 \times
  10^{10}\mathrm{g\ s^{-1}}$. C, I, W and $\Delta t$ are defined as
  in Table \ref{tab2}. }
\label{tab3}
\end{table*}

Figs \ref{fig3}(a)-(c) show the density distribution of the absorbing
material on the orbital plane and for the same cases as those in Figs
\ref{fig2}(a)-(c), revealing some small changes of the absorbing areas
in line with the discussion given above.

By comparing the isotropic models with the observations it is possible
to estimate  ${\dot M}_p$ that match the proposed observational
absorption ranges for a given $T_{0}$ and $v_{o} $ pair.  Relatively
low coronal temperature values ($T_{0}\approx 3$ MK)
together with a range of $v_{o}\sim (250 - 430)$ km s$^{-1}$
values (e.g. models a and b of Table \ref{tab2}) are associated with
${\dot M}_p \approx 3 \times 10^{10}$ g s$^{-1}$
independently of the heat transport efficiency ($\Gamma \sim
[1.01-1.13]$).  Intermediate and high values of the coronal
temperatures ($T_{0}\approx (4-7)$ MK) together with a broad
range of $v_{o}\sim (400 - 850)\mathrm{km\ s^{-1}}$ values lead to the
range ${\dot M}_p \approx (3-5) \times 10^{10} \mathrm{g\ s^{-1}}$,
depending upon the efficiency of the stellar wind to transport heat
($\Gamma \sim [1.01-1.13]$, upper and bottom panels of Table
\ref{tab2}, respectively).  The Ly$\alpha$ transit time variation
obtained while increasing the coronal temperature varies from $\Delta
t \sim 4.4$h to $\Delta t \sim 5.5$h. Note from fig. 3 in VM03, that the observational Ly$\alpha$ transit time
can be estimated as $>4$ h.

\begin{figure*}
\centering
\includegraphics[width=10cm]{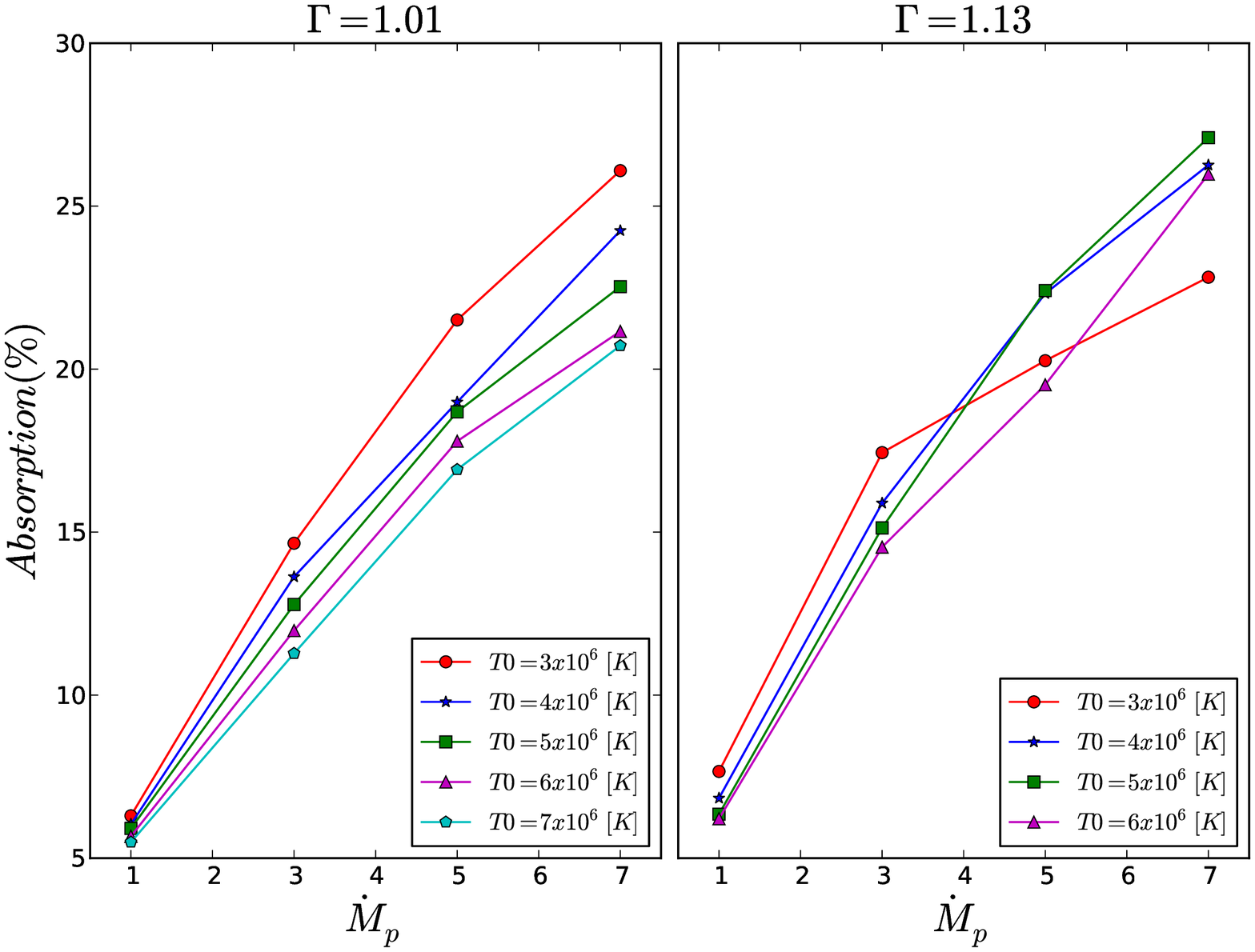}
    \caption{Maximum Ly$\alpha$ absorption during transit as a function of the
      planetary mass-loss rate for the isotropic models (considering the line core limits, C: [19-11] per cent) with different T$_0$ for $\Gamma= 1.01$ (left-hand panel) and $\Gamma= 1.13$ (right-hand panel). }
    \label{fig7a}
 \end{figure*}

Finally, in Fig. \ref{fig7a} it is
possible to see a comparison among the isotropic models,  $\Gamma= 1.01$ (left-hand panel),
and $\Gamma= 1.13$ (right-hand panel) of the maximum absorption value during transit as a
function of the planetary mass-loss rate considering the line core limits C: [19-11] per cent. 
From the left-hand panel, we see that models
with $\Gamma=1.01$ emission slowly begin to reach saturation, whereas, as expected,
signs of saturations are already met in the case with $\Gamma=1.13$
and stellar wind temperature ($T_0=3$ MK),  resulting in the intersection of some of the lines.
 This is the reason why, unlike the general trend mentioned earlier, we find an 
increase of the absorption with increasing coronal
temperature.  That is,  the bottom panel  of Table \ref{tab2} shows that, when varying $T_{0}$ from $3$ 
to  $ 5 $ MK with  ${\dot M}_p=5 \times 10^{10}$g s$^{-1}$
 the absorption values increase from $20.26$ to $22.41$.
 
It is worthwhile to notice that all these results are valid considering a fix value of  ${\dot M}_{*}$.
In order to test the influence of ${\dot M}_{*}$ in the calculated Ly$\alpha$ planetary absorption  we run two cases where  ${\dot M}_{*}$ is $0.1$ and $10$ of its fixed value ($\dot{M}_{\star}=9.0 \times 10^{-14}M_{\odot}$~yr$^{-1}$).  The other model parameters are: $T_{0}=4$ MK, $\Gamma= 1.13$ and  ${\dot M}_p=3 \times 10^{10} \mathrm{g\ s^{-1}}$. We found that the planetary absorption increases almost a $40\%$ in the first case and diminishes by the same amount  in the second case. This 
implies that the  ${\dot M}_p$ that better adjust the observational Ly$\alpha$ absorption values  are  ${\dot M}_p=1 \times 10^{10}$ and  $5 \times 10^{10} $g s$^{-1}$, respectively.

\subsection{Anisotropic planetary wind models}
 
\begin{figure*}
\centering
\includegraphics[width=18cm]{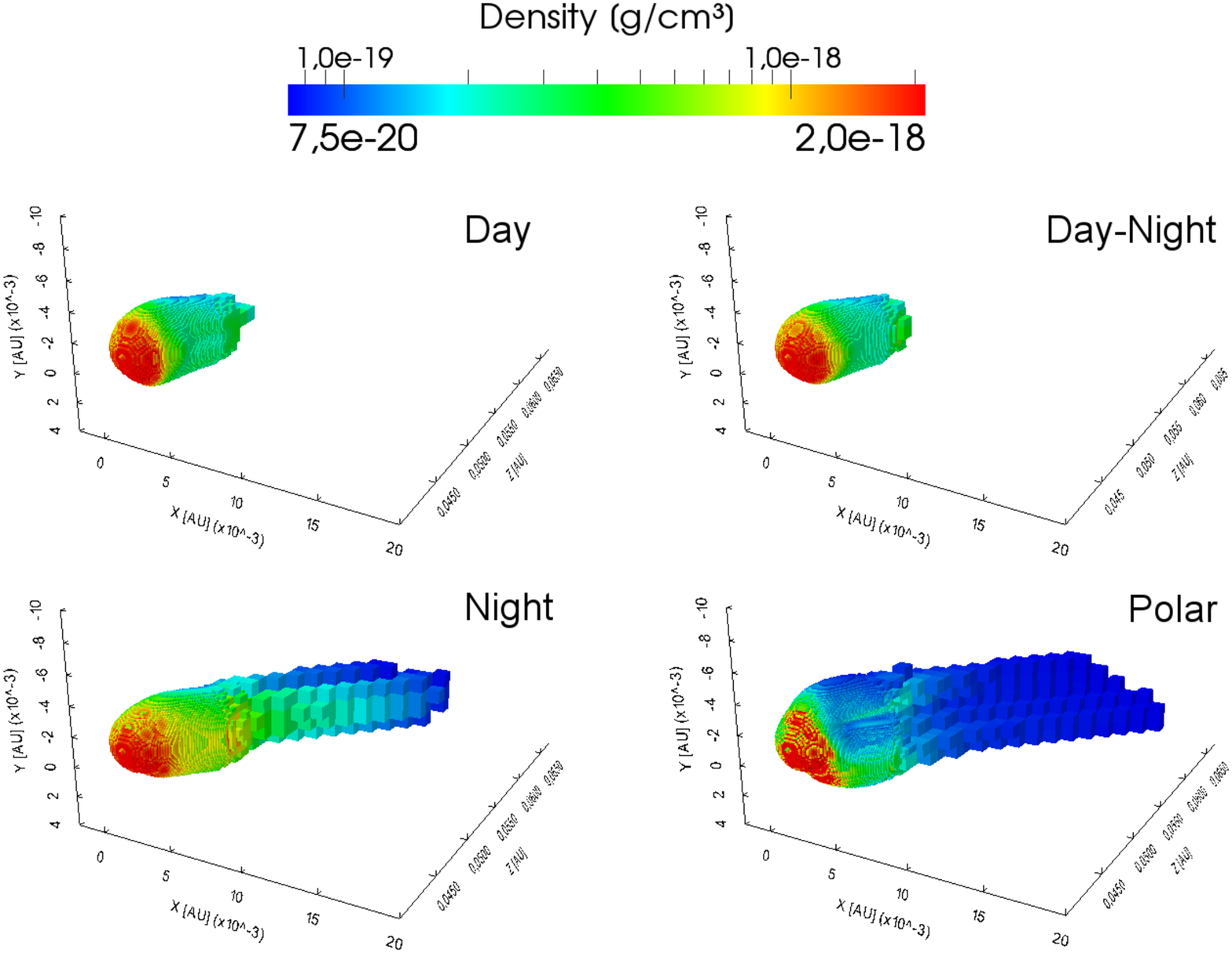}
    \caption{3D surface density distribution of the 
      absorbing material corresponding to the anisotropic planetary wind models,
      obtained for $\Gamma = 1.13 $, ${\dot M}_p = 3 \times
      10^{10}$ g s$^{-1}$ and stellar wind temperature of $T=4$ MK. Upper left: day; upper right: day-night; lower left: night and lower right: polar. All figures are obtained when the planet is located at the same  orbital position.}
    \label{fig6}
 \end{figure*}

In Section \ref{toy} we introduced four anisotropic models inspired on the
discussions in \citet{adams,gar07,murr} and \citet{yelle2004}.  Table
\ref{tab3} shows the results obtained for the different planetary wind
models ordered with decreasing absorption (see columns 2-4).  From the
results, we can see that the Day and Day-Night anisotropic models
produce less absorption and a shorter transit time than the isotropic
one. Plasma mostly evaporated from the illuminated side is rapidly
ionized and thus absorption is less important than in the other escape
models.  The shortest transit time is produced when the planetary wind
is emitted from the Day-Night model. 
 The difference between  this last 
model, with respect to the Day one, is because 
there is less material to be ionized given the lower initial planetary
speed value (to allow the same ${\dot M}_p$).
The
largest absorptions and transit times occur for the Polar and Night
models. As can bee seen in figure \ref{fig6}, where a 3D view shows the density distribution for the absorbing
material, the Day and Day-Night models exhibit smaller absorbing
tails, and the longest belongs to the Polar model. This latter model
exhibits greater neutral density (in comparison to the rest of the
anisotropic models) available to absorb within the tail, and can
propagate further before being heated and consequently ionized.

\begin{figure*}
\centering
\includegraphics[width=\textwidth]{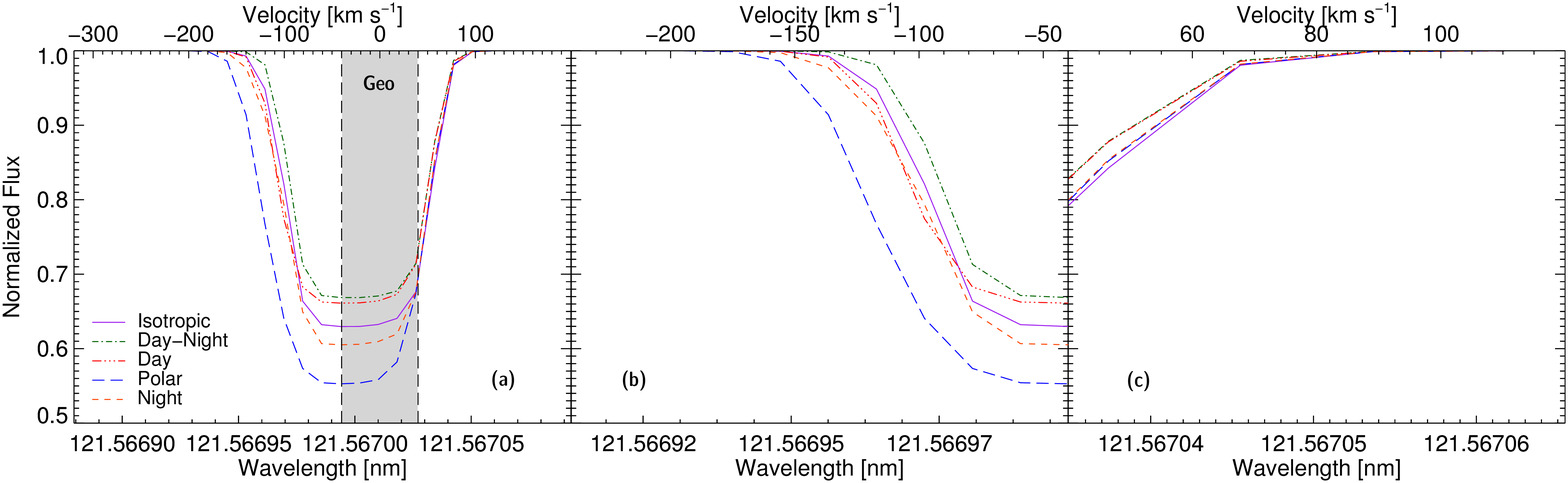}
    \caption{$Ly\alpha$ normalized flux (to the out of transit value and for the C range) as a function of the wavelength for the anisotropic models with $\dot M=3 \times 10^{10}$ g s$^{-1}$ and
      T$=4 $ MK, and the corresponding isotropic model. The
      grey stripe represents the excluded area due to geocoronal
      contamination: panel (a) shows a comparison among models with different
      planetary atmospheric structures as depicted in Fig.
      \ref{fig1}. Panels (b) and (c) show a zoom of the right- and left-hand wings
      for each case. All figures are obtained when the planet is located at the same  orbital position. }
      \label{fig7}
\end{figure*}

Fig. \ref{fig7}(a) shows the $Ly\alpha$ simulated normalized flux (to the out of transit value for the C limits) as a function of the wavelength for
the same anisotropic models ($\dot M=3 \times 10^{10}$ g s$^{-1}$ and
T$=4 $ MK), together with the corresponding isotropic model. 
Figs \ref{fig7}(b) and (c) show a zoom of the right- and left-hand wings of the
absorption profile as a function of the wavelength for the previous
mentioned cases. The evident blueshifted asymmetry in the line
profiles is a consequence of the absorbing planetary material escaping
towards the line-of-sight direction (VM03), being more
pronounced in the Polar and Night cases, where the planetary
mass-loss configuration favours the driving of absorbing material
towards the line of sight. 
\\

\section{Conclusion}
\label{conclusion}

In this work, we have explored a wide range of  parameters with the
main purpose of studying the dynamic response of close-in exoplanets
to different stellar wind conditions. The chosen system, HD 209458 is
the most known among the many observed, and therefore has served as a
good numerical laboratory. Initially, our aim was to find a mechanism
that would help to determine the stellar wind velocity at the orbital
position, yet not defined for the present system due to the need of
better observational data. Nevertheless, from our results we have been
able to narrow down some parameters.
For a fixed ${\dot M}_\star$, we found that the ${\dot M}_p$ is not too
sensitive to the stellar wind conditions, in comparison with the
ranges reported in the literature. Observation of transits have found
mass-loss rates in the range $10^9-10^{11}$ g s$^{-1}$ (see VM03; \citealp{lecavelier2010}; 
\citealp{linsky2010} and \citealp{guo}).
 Assuming  ${\dot M}_p$ values varying in a  reasonable range ($\dot M_p =$  [1-7] $\times 10^{10}$g
s$^{-1}$) we could constrain  the planetary mass-loss rate by comparing the observational Ly$\alpha$ absorption with the numerically derived values.
The range we found for $\dot M_p$ is $\sim$[3-5]$\times
10^{10}$ g s$^{-1}$, where it changes a factor of $\sim 1.7$
for a change in temperature from [3-7] $\times 10^6$ K, a variation of
the stellar wind velocity in the range $\sim[250-850]$ km s$^{-1}$,
and a polytropic index value between $\Gamma \sim [1.01-1.13]$. Also,
our models give a variation of the transit time between $[4.4-5.5]$h,
which is in agreement with the minimum observed $\Delta t$ (see fig. 3 by VM03 where we can estimate a
$\Delta t > 4$ h).
A better observational determination of $\Delta t$ could play an
important role in solving part of the degeneracy in the inferred mass
loss rate and stellar wind parameters. 
The Ly$\alpha$ planetary absorption depends on the  ${\dot M}_{*}$ value,  thus assuming that a fix value of this parameter    constraints the results. However, by studying a single model, we notice that while varying ${\dot M}_{*}$ through a factor of $0.1$ and $10$  the ${\dot M}_{p}$  that better adjusts the observational Ly$\alpha$ absorption value changes 
in a factor of $0.3$ and $1.7$, respectively.

Another important result comes from studying the variation of the
structure of the escaping atmosphere. Even though the present models
are far from being realistic, they give us a hint on how to further
study these type of systems. Still, we can conclude that there is a
noticeable effect produced by changes in the escaping atmosphere on
the observed Ly$\alpha$ absorption. The Polar and Night side emitting
models, show an enhanced absorption with respect to the isotropic
and there is an appreciable
reduction on the maximum absorption measured for the Day and
Day-Night cases, where the Day-Night model shows the minor value.
 Note that for all isotropic and anisotropic models the velocity of the planetary escaping material ranges from 
 $\sim [-160, 100]$ km s$^{-1}$, in accordance with the observational data. 

More realistic models are necessary, specially if we want to venture
on this last part. For instance, an improved model should include
explicitly the photoionization of the planetary wind in the
hydrodynamical calculation, even though, this is unlikely to produce
an important effect on the Ly${\alpha}$ absorption, because as the
tail expands it becomes optically thin.  Also, a similar study to the
present one should be done with 
the inclusion of the systems magnetic
fields, i.e. the almost certain planetary magnetic field and the
stellar magnetic field is another.

\bigskip

PV, AR and AE acknowledge financial support from Conacyt grant 167611
and DGAPA-UNAM IN105312. Also, all authors acknowledge financial support from 
Conacyt-CONICET grant CAR 190489.


\end{document}